\documentclass[aps,prl,twocolumn,showpacs,floatfix,preprintnumbers,amsmath,amssymb]{revtex4}
\usepackage{epsfig}
\usepackage{graphicx}
\usepackage{dcolumn}
\usepackage{bm}

\begin{document}
\title{Quantised orbital alignment and Identical High-K Bands in A=180 region}
\author{Aswini  Kumar  Rath$^{1}$, Z Naik$^2$, C R Praharaj$^{3}$}
\affiliation{$^1$ P.G. Department of Physics, Sambalpur University, Burla, 768 
019, INDIA \\
$^2$ Tata Institute of Fundamental Research, Mumbai, 400005, India\\
$^3$ Institute of Physics, Bhubaneswar-751 005,India} 
\begin{abstract}
The origin of the identical gamma-ray energies (IB)  observed in the
K=33/2$^+$ and K=16$^+$ bands in  $^{177}$Lu and  $^{178}$Hf respectively
is investigated using deformed Hartree-Fock and Angular Momentum Projection technique.
We find that quantised orbital (rather than spin) alignment of protons in the
m=1/2$^+$ inert orbit leads to identical energy spectra.
A K=16$^+$ band with identical spectrum is predicted in  $^{176}$Yb at 4.48 Mev
of excitation.
\end{abstract}
\maketitle

The fascinating phenomena of Identical Band (IB)  was first observed in the
superdeformed (SD) bands of neighbouring even-even ($^{152}$Dy) and odd-even ($^{151}$Tb)
nuclei during 1990 [1,2]. The $\gamma$-ray energies of the SD bands of $^{152}$Dy and
$^{151}$Tb were found to be identical within 0.01\% . Soon this phenomena
was identified in the normally deformed even-even
and odd-A rare-earth nuclei . The recent observation [3] of this identity in the
 bands based on high-K isomeric states in the neighbouring nuclei ($^{178}$Hf and
$^{177}$Lu) is interesting. The transition energies observed for a few states in the 
K=33/2$^+$ (T$_{1/2}$=902s) band in $^{177}$Lu, have exactly the same value as of the
K=16$^+$ (T$_{1/2}$=31y) band in $^{178}_{ 72}$Hf. 
 These high-K states occuring at a few MeV of excitation  having four-five unpaired
nucleons 
are isomeric and have intermediate deformation {$\beta$= 0.28} 
. Compared to SD states the configurations of high-K states are 
fairly well known and hence study of IB in high-K states can enhance our understanding
of the phenomena.

Explanation of this phenomena has been tried
in various models [2,4-6]. These models invoke the phenomena of pairing, quantised spin
alignment, pseudo-spin symmetry, supersymmetry etc. in nuclei.
Quantised spin alignment leading to identical superdeformed bands was
reported by Stephens et-al [4]. Cheng-Li Wu et-al [5] have conjectured that 
the quantised spin alignment may be model dependent. However the mechanism of
identical high-K bands have not been identified. Using a quantum
many body method based on deformed Hartreee-Fock and angular momentum
projection we find that identical high-K bands in $^{178}$Hf and $^{177}$Lu
 occur due to quantised orbital 
alignment of protons in the inert orbit. The
relative orbital aligment between the J+${1\over2}$ states in the K=33/2$^+$
band in $^{177}$Lu and J states in K=16$^+$ bands of $^{178}$Hf and $^{176}$Yb
 are exactly ${1\over2}\hbar$. The relative spin alignment between these states
are negligible.

The model used by us is  based on a quantum many-body method
 which has been quite successful in
explaining the high-spin spectroscopy in A=180 region [7,13] and light
mass region [8] also.
It is based on deformed Hartree-Fock model for the intrinsic states and
Angular Momentum Projection (or J projection, for short) for the physical
states based on these intrinsic states.
\begin{figure}[htbp]
\begin{center}
\includegraphics[width=8.0cm,height=6.0cm]{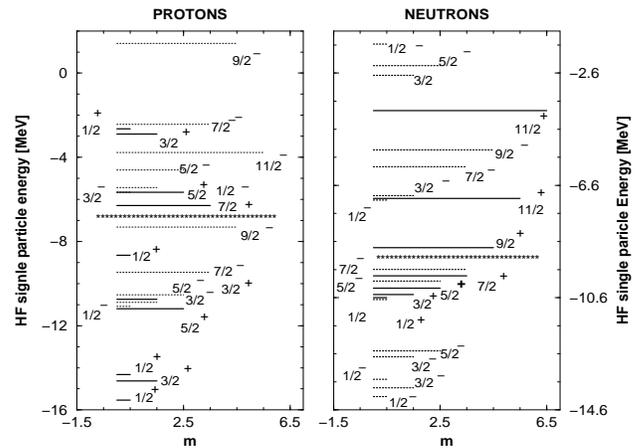}
\end{center}
\caption{The prolate Hartree-Fock single-particle orbits 
 for protons and neutrons  are shown for  $^{178}$Hf.  
  Solid and dashed lines correspond to the positive and negative parity orbits, 
  respectively. 
The length of these lines indicate the magnitude
of the z-projection (m) of the angular momentum.
Stars guide the eye to the Fermi level. The inert m=1/2$^+$ proton 
orbit is just below the m=9/2$^-$ 
Fermi level and the deformed m=1/2$^+$ proton orbit is third from the bottom.
}
\end{figure}

In deformed (axial) Hartree-Fock and 
angular momentum projection (PHF) technique [for details see [8,9]  and references there in]
we start with a model space 
and an effective interaction. The model space is presently
limited to one major shell  for protons and neutrons
lying outside the $^{132}$Sn core. 
Reasonable spherical  single particle energies [10] and surface-delta interaction 
(strength has been taken to be 0.3 MeV for pp, nn and pn interactions
respectively) are used in our calculation.
 The 
3s$_{1/2}$, 2d$_{3/2}$, 2d$_{5/2}$,  1g$_{7/2}$, 1h$_{11/2}$ and
1h$_{9/2}$ 
 proton 
states have energies  3.654, 3.288 , 0.731, 0.0, 1.705 and 7.1  MeV, and the 
3p$_{1/2},$ 3p$_{3/2}$, 2f$_{5/2},$ 2f$_{7/2}$, 1h$_{9/2}$ and 1i$_{13/2}$  
neutron states have energies 4.462,   2.974,  3.432,  0.0,  0.686 and
1.487 MeV  respectively.
The prolate HF calculation for the valence nucleons 
lying outside the $^{132}$Sn core is 
performed for both the nuclei.
The time reversal symmetry preserved for the even-even
$^{178}$Hf nuclei for the K=0$^+$ configuration
is broken  slightly in the
odd-A case $^{177}$Lu (where the $\pm$m orbits are no more degenerate). 
 However, for the high-K states
it has been shown [11] that this symmetry is badly broken.
 The 
set of  prolate deformed HF  orbits  (with well defined
m-quantum numbers) shown in Fig.1 
for $^{178}$Hf forms the deformed single particle basis for the valence protons and
neutrons.
This basis is  enriched compared to the
Nilsson basis as  the pp,nn and pn correlations are built in by the
inclusion of  residual
interaction in a self-consistent manner through the HF iteration  procedure.
Occupation of the lowest HF orbits by the active neutrons and protons
forms the ground band (K=0) intrinsic configuration with a well defined
K  quantum number. This intrinsic HF wavefunction $\mid\Phi_K
\rangle$ is
a superposition of states of good angular momenta which are projected using the
angular momentum projection operator:

\begin{equation}
P_{K}^{IM}={{2I+1}\over {8 \pi^{2}}}\int d \Omega {D_{MK}^{I}}^{*}(\Omega)\,
R(\Omega)\,
\end{equation}
The angular-momentum-projected normalised states are given by\,
\begin{equation}
{\Psi_{K}^{IM}} ={ {P_{K}^{IM} \mid  \Phi \rangle} \over {\sqrt{\,
\langle  \Phi \mid P_{K}^{IK}\,
\mid \Phi \rangle}}}\,
\end{equation}

 The energy of the states are obtained from the hamiltonian overlap  given by,
\begin{eqnarray}
&&\langle \Psi ^{I}_{K_2} \mid H \mid \Psi\,
^{I}_{K_1}\rangle\nonumber\\
&=& {{2I+1}\over{2}}\,
\frac{1}{(N^{I}_{K_{1}K_{1}}N^{I}_{K_{2}K_{2}})^{1/2}}\nonumber\\
&& \times\int d\theta \sin \theta d^{I}_{K_{2}K_{1}}(\theta)
 \langle \phi_{K_{2}}  | H  e^{-i \theta J_{y}} |
\phi_{K_{1}}\rangle \,
\end{eqnarray}
with $N^{I}_{KK}=  \langle  \Phi  \mid P_{K}^{IK}  \mid  \Phi \rangle$ .
 $N^{I}_{KK}$ essentially represents  the intensity of angularmomentum I
in a K configuration. Interestingly N$^{I}_{KK}$ for the
I=16,17,18 ... states of
K=$16^+$ band in $^{178}$Hf 
 is found to be identical respectively with those of
the I=33/2,35/2,37/2... states of the K=33/2$^+$ band in $^{177}$Lu.

Using these projected states the expectation values of various
tensor operators including E2,M1, $\vec J$, and $\vec L$ are evaluated.
 Evaluation of  the matrix elements of the opertaor J  for the 
projected states  provide the information about the nature of alignment of
the
protons (neutrons) in individual orbits for the J (or I) states.
Similarly   calulation of $\langle JJ\mid L_z\mid JJ\rangle$
and $\langle JJ\mid S_z \mid JJ\rangle$ provide the
spin and orbital alignments of nucleons in various shell model orbits 
in a K configuration (see Fig.3).
This provides a microscopic explanation of the
identical  high-K bands.
 
Dracoulis et-al [3] have identified
the IB in K=33/2$^+$ band in $^{177}$Lu and K=16$^+$ in
$^{178}$Hf at excitation of 2.771 and 2.447 MeV respectively.
 The excitation energies of these high-K IB in these nuclei agree
fairly  well with our calculated values of 3.252 and 2.255 irespectively. 
The  K=16$^+$ and the K=33/2$^+$ isomersi are obtained by
particle hole excitations over the ground state have the following intrinsic
structures.
 K=16$^+$ in $^{178}$Hf - (7/2$^+$9/2$^-$)$^p$ and (7/2$^-$9/2$^+$)$^n$.\hfill
 K=33/2$^+$ in $^{177}$Lu -  (7/2$^+$9/2$^-$1/2$^+$)$^p_{inert}$ and (7/2$^-$9/2$^+$)$^n$.\hfill
 J projection ( 
AMP for short) from these high-K structures
 give rise to the identical high-K bands
(see Table-1 and Fig.2).
 The $\gamma$-ray transition energies
of the high-K states
 are found to be identical 
within 10-20 keV. The absolute values of the theoretical (before mixing)
 $\gamma$-ray energies are overestimated compared
to the experiment.  Band mixing 
 improves the agreeement considerably (see Table).  Similar High-K IB in neighbouring ($^{179}$Lu,$^{180}$Hf) and ($^{181}$Lu,$^{182}$Hf)  
nuclei are predicted.[12]
\begin{table}
\caption{ The $\gamma$ ray energies (E$_\gamma$=E$_I$-E$_{I-1}$) of the high-K bands are compared with
experiment(EXP) . The agreement improves after band mixing (BM). 
IB for $^{176}$Yb is predicted.}
\begin{tabular}{|l|l|l|l|l|l|l|l|l|l|}
\hline
$^{178}$Hf&{PHF}&{BM}&EXP&$^{177}$Lu&PHF&BM&EXP&$^{176}$Yb&PHF\\
K=&&&&K=&&&K=&K=&\\
16$^+$&&E$_\gamma$&&${{33}\over{2}}^+$&E$_\gamma$&&16$^+$&16$^+$&E$_{\gamma}$\\
I($\hbar$)&keV&keV&keV&I($\hbar$)&keV&keV&keV&I($\hbar$)&keV \\
\hline
17&579&419&357&35/2&591&443&357&17&580 \\
18&613&464&378&37/2&624&487&377&18&613 \\
19&646&506&398&39/2&657&528&&19&646 \\
20&679&545&417&41/2&689&566&&20&679 \\
21&712&582&436&43/2&721&602&&21&712 \\
22&744&617&454&45/2&752&637&&22&744 \\
23&775&652&474&47/2&783&669&&23&775 \\
24&806&685&&49/2&814&701&&24&806 \\
25&836&716&&51/2&843&732&&25&835 \\
26&865&747&&53/2&871&762&&26&865 \\
27&893&777&&55/2&898&791&&27&892 \\
28&919&806&&57/2&923&818&&28&919 \\
29&943&834&&59/2&945&847&&29&946 \\
30&966&860&&61/2&965&875&&30&970 \\
\hline
\end{tabular}
\end{table}

Let us analyse the miscroscopic  structure of the
intrinsic HF configurations of the high-K isomers.
The two high-K structures differ in occupation of 
  proton in  the m=1/2 orbit.
 Hence the role of
this  unpaired proton in the natural parity orbit ($m$=1/2$^+$) in
K=33/2$^+$ of $^{177}$Lu becomes important in understanding the IB.
Let us examine the HF wave function of this orbit.

{\it The Role of the natural  parity HF inert orbit m=1/2$^+$}:
The HF wave function ( -.388$\mid 3S_{1/2}\rangle$+.632 $\mid 2d_{3/2}\rangle$+.438 $\mid 2d_{5/2}\rangle$
-.508 $\mid 1g_{7/2}\rangle$)
 of this orbit indicate considerable mixing from all
the four shell model states with $d_{3/2}$
dominance.
 The mixing is such that
the quadrupole deformation of this orbit is
  negligible (i.e  -0.009 b$^2$, where
b is the harmonic oscillator length parameter). Hence 
occupation or non occupation of this orbit does not change the deformation and 
the moment of inertia significantly.
So the transition energies of the high-K band in $^{176}$Yb
 where the inert orbit is empty
should  also be identical to the K=16$^+$ band of $^{178}$Hf 
and K=33/2$^+$ band of $^{177}$Lu.
In fact (see fig.2 ) we find that the spectrum of the K=16$^+$ structure
 in $^{176}$Yb is identical with the K=16$^+$ band
 of $^{178}$Hf and K=33/2$^+$ band of $^{177}$Lu.
 But if a deformed m=1/2$^+$ orbit is occupied
(as in case of K=33/2$^+)_{def}$ of $^{177}$Lu in fig.2) the spectra is not identical
to the K=16$^+$ band in $^{178}$Hf. It is
because this m=1/2$^+$  proton HF orbit ($3^{rd}$ from the bottom of Fig.1) is not inert and contributes to deformation
 (i.e quadrupole moment of this
deformed orbit is 1.341 b$^2$).
 It appears that the pair
 of protons in the
well mixed m=1/2$^+$ natural parity inert orbit remain as spectator in
 $^{178}$Hf.
In $^{177}$Lu the lone proton aligns its angular momentum with $^{176}$Yb as
the rotating core. Experimental observation of this K=16$^+$ band in $^{176}$Yb
predicted to be at about 4.48 MeV of excitation in our calculation
is essential to confirm the above proposition.
It is clear that occupation or non ccupation of this orbit
matters little to the energy spectra. Thus this orbit really acts as an
inert orbit (as evident from its contribution to quadrupole moment)
 and leads to identical band.

\begin{figure}[htbp]
\begin{center}
\epsfig{file=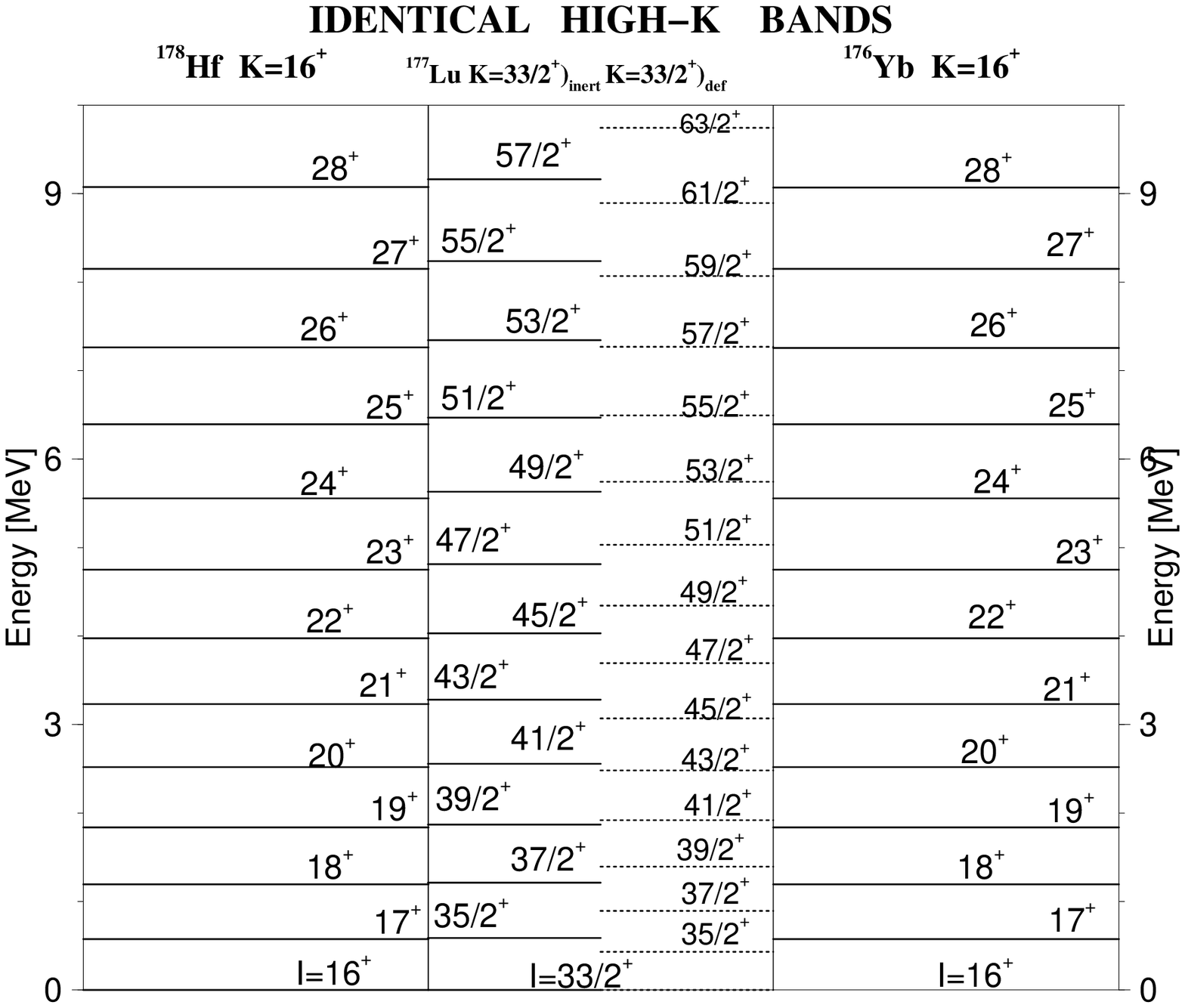, height=6.0 cm, width=8.0 cm}
\end{center}
\caption { 
The energy spacings of two 
 K=33/2$^+$ bands
(one with deformed m=1/2$^+$ proton orbit and the other
with an {\bf inert} m=1/2$^+$ orbit)  
 in $^{177}$Lu are
compared with those of 
K=16$^+$ (in $^{178}$Hf). The K=33/2$^+$)$_{inert}$  
band is identical to K=16$^+$ band in $^{178}$Hf.
and  $^{176}$Yb.}
\end{figure}

Since the neutron configuration is same for both K=16$^+$ 
and K=33/2$^+$ bands , the role of neutrons must be identical in these
two bands. It is found that the amount of J carried by the neutrons in 
various  shell model orbits included in the model space
are identical for both these bands. 
Except for the 1g$_{7/2}$ and 2d$_{5/2}$ protons the
 contribution to angular momentum  from the
rest of the protons and neutrons  in various other orbits are same for both
the nuclei. 
\begin{figure}
\begin{center}
\epsfig{file=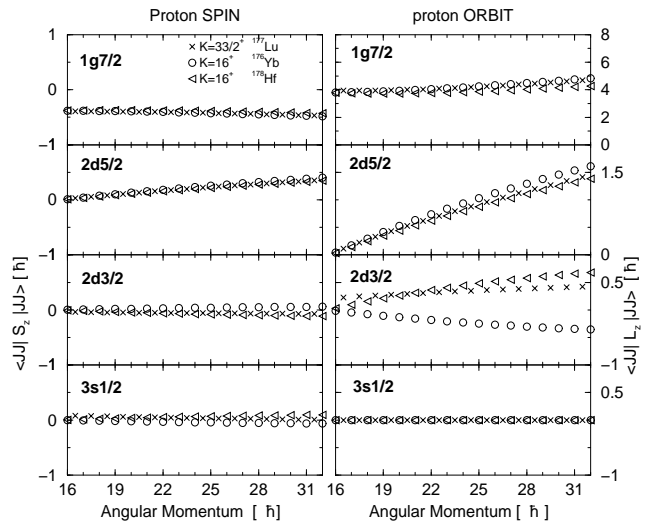, height=7.0 cm, width=8.5 cm}
\end{center}
\caption {
The  orbital and spin alignment from
various  proton orbits are compared with 
K=16$^+$ (in $^{178}$Hf, $^{176}$Yb) and K=33/2$^+$
bands (in $^{177}$Lu).
}
\end{figure}

As emphasized before the two high-K configurations
differ in the occupation of protons in the inert orbit m=1/2$^+$. Hence the
angular momentum carried by  protons in these natural parity orbits needs 
microscopic analysis to pin point the underlying reason
for identical energy spacing. In Figure 3
both the orbital and spin angular momenta contributions from individual
proton orbits are shown. As evident from the figure the
spin contributions from various proton orbits are same for
$^{178}$Hf, $^{177}$Lu and $^{176}$Yb but the orbital
  contribution from
orbits (other than 3s$_{1/2}$) are not identical. We
sum these contributions from different orbits both for
orbital and spin parts
and substract the contribution of $^{178}$Hf from $^{177}$Lu
(as shown in upper most panel of Fig. 4). One finds that the relative orbital alignment
is quantised (L$_{diff}$=1/2$\hbar$). 
The relative spin alignment (S$_{diff}$) is almost zero for
the identical bands. Similar plots for the case of K=33/2$^+)_{def}$ in $^{177}$Lu
shown in the lower pannel of the Fig.4 . As is evident the orbital
contribution slowly increases
for the deformed case  and saturates to 1$\hbar$ at higher spin.
Although spin contribution is non-zero it remains constant (and hence quantised
) for all J.
Infact quantised spin alignment of 1/2$\hbar$ have been reported for the
identical superdeformed bands  by Stephens et-al.[4]
But we find the spin alignment to be negligible in these two high-K IBs. 
\begin{figure}[htbp]
\begin{center}
\epsfig{file=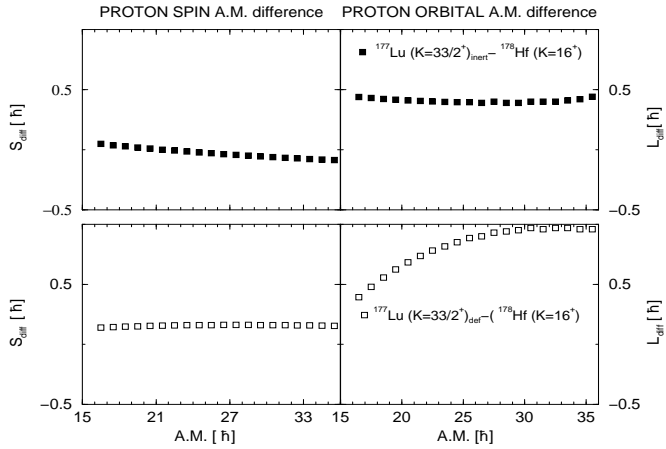, height=6.0 cm, width=8.5 cm}
\end{center}
\caption { 
The protn orbital and spin alignments in
 $^{177}$Lu (K=33/2$^+$) relative  $^{178}$Hf (K=16$^+$)
are given for the inert case (upper pannel)
as well as deformed case (lower panel). As evident the orbital angular momentum
is quantised for the inert case (i.e identical bands) and not so for the
deformed case ( non-identical
bands).}
\end{figure}

The PHF calculation  well reproduces the lowlying band structures
in $^{178}$Hf and $^{177}$Lu nuclei. 
High-spin states in several bands are also predicted.
Our calculation
 reproduces the identical high-K bands in these nuclei.
The  K=16$^+$ band,  predicted in $^{176}$Yb,  
is found to be identical  with the K=16$^+$ band (of $^{178}$Hf) and K=33/2$^+$ 
band (of $^{177}$Lu). This identity suggests that $^{176}$Yb acts as a core
in the high-K isomers of $^{177}$Lu and $^{178}$Hf with the rest of the 
protons being spectator.  We find that 
quantised orbital alignment of protons in the m=1/2 natural parity inert orbit
leads to identical high-K bands in neighbouring nuclei. Thus 
 the inert orbit near the
Fermi level leads to quantised "orbital alignment" of nucleons in this orbit.
The deformed configuration mixing
 leads to the  inertness of the  m=1/2  proton 
orbit, the occupation/non-occupation of which
does not affect the energy spectra. The variation in the
nature of orbital/spin alignments in the identical bands starting 
from normal deformed to superdeformed nuclei need further investigation.

One (AKR) of us  thanks Institute of Physics, Bubaneswar for
library and computer facilities for this work during
summer visit.
Thanks are due to the DST, BRNS government of India for  financial support. Last but not the
least our sincere thanks to Prof. S P Pandya
for suggesting us  to work on the IB in A=180 region.

\end{document}